# Heteromated Decision-Making: Integrating Socially Assistive Robots in Care Relationships

Richard Paluch

Information Systems/IT for the Ageing Society, University of Siegen, Germany, richard.paluch@uni-siegen.de

Tanja Aal

Information Systems/IT for the Ageing Society, University of Siegen, tanja.ertl@uni-siegen.de

Katerina Cerna

Intelligent Systems and Digital Design, School of Information Technology, Halmstad University, Sweden,

katerina.cerna86@gmail.com

Dave Randall

Information Systems and New Media, University of Siegen, Germany, dave.randall@uni-siegen.de

Claudia Müller

Information Systems/IT for the Ageing Society, University of Siegen, Germany, claudia.mueller@uni-siegen.de

Technological development continues to advance, with consequences for the use of robots in health care. For this reason, this workshop contribution aims at consideration of how socially assistive robots can be integrated into care and what tasks they can take on. This also touches on the degree of autonomy of these robots and the balance of decision support and decision making in different situations. We want to show that decision making by robots is mediated by the balance between autonomy and safety. Our results are based on Design Fiction and Zine-Making workshops we conducted with scientific experts. Ultimately, we show that robots' actions take place in social groups. A robot does not typically decide alone, but its decision-making is embedded in group processes. The concept of heteromation, which describes the interconnection of human and machine actions, offers fruitful possibilities for exploring how robots can be integrated into caring relationships.

## 1 INTRODUCTION

Robots are already being used in a range of settings and are presumed to be capable of making important decisions. Often the tasks taken on are job-specific, and what has long been commonplace with regard to industry and manufacturing has now found its way into care. In light of demographic change and an aging society, it is now possible to see how robots might be used to support care practices [1,6,8,24,62]. In this context, these robots are referred to as socially assistive robots (SARs) and are primarily intended to support interaction-based care rather than simply performing mundane functions. Complementing caregivers is still, however, something of an unrealized ambition and even with technical feasibility it remains unclear whether such changes constitute unalloyed benefit. Certainly, technocentric assumptions have been subject to some critical scrutiny [3,51].

In the research discourse, autonomy and safety are of great importance for the design and understanding of Human-Robot Interaction (HRI) and decision-making (e.g., for an overview, see [36]). Furthermore, many nursing fields also address the importance of autonomy and safety, for example, when caring for persons with dementia [37,38,54,61]. In the case of persons with dementia, there is a negotiation regarding how much autonomy they might be allowed while maintaining safety. Balancing autonomy and safety is a highly contextual matter here. It depends on institutional and domestic circumstances, the degree of vulnerability of the care receiver, ethical considerations, and many more aspects. This is similar for robot use. In this context, more autonomy always means a certain safety risk. In this respect, it can be seen that the question of safety and autonomy touches on different, social, ethical and legal aspects that are of great significance in an aging society [25].

In this contribution, we focus on the broad question which emerged in our work: "When designing for interactions with robots, how might we balance autonomy and safety in group-robot interaction?" Our guiding question is not exclusively about technical feasibility. We understand design here not as the technical development of a product, but as an innovative solution for helping people now and in the future to reduce their vulnerability [29,35,55,63]. In addition, autonomy and safety are interactional and mediated in caregiving situations by caregivers and care recipients. Our approach is redolent of the "group turn" that can currently be observed in HRI [46]. This is not an entirely new finding [27,45,48,56], but it is gaining more and more attention in the current discourse [2,12,17,18,22,23,39,60]. This is also in line with HCI research perspectives [26] which is why we refer to group constellations when we focus on decision-making of robots.

In this way, a vision can be developed of what role future technologies such as robots are supposed to play in our society and, to address the CHI workshop question, how to integrate SARs "into existing relationships in assistive scenarios to build trust with users?" For this, there must be a balance between autonomy and safety or between decision-making and trust. That can be achieved through heteromated decision-making. That is, one in which robots and humans both mediate the decision-making process [14].

## 2  STATE OF THE ART

With regard to decision-making the concept of autonomy is frequently used in the literature on robots. Different aspects are identified by Bekey [5], who states that autonomy can be present in a technical system if it can operate in a real environment for a certain time without being externally controlled (see also [10]).

The extent to which a technical system can be completely autonomous or indeed whether this can even be planned for, remains an open question. For example, Castro-González et al. [10] claim that it is completely impossible for a robotic system to be fully autonomous. In the design process, there is already the idea of what tasks robots should be used for. Their situatedness in the world is thus predetermined and mediates their interaction with the environment. Having said that, the degree of autonomy (or agency) that can be ascribed to human beings is also debatable. Humans are to a certain extent dependent on their fellow human beings and in the environment in which they find themselves. Indeed, the structure/agency debate is one of the cornerstones of the sociological problematic (see e.g. [20]). It has also been extensively discussed in the guise of 'cultural robotics' [47].

Nevertheless, robots are also designed to have at least some limited agency insofar as they can be programmed to make decisions [21]. Here, autonomy is determined by how many decisions they can make [19,43,44] or how mobile they are in an environment [53]. Regarding HRI, studies have been conducted on how robots can become more autonomous [41,42]. However, robots are still far from reaching the same level as humans. Their autonomy and mobility is mostly possible in highly controllable environments only [28]. Lindemann and Matsuzaki [32] describe, based on their ethnographic data, the extent to which the autonomy of robots can be described in terms of temporal and spatial aspects. To simplify, robots perform actions in a digital space and time that is computable. Any spatial point in which an object is located can be expressed in terms of numbers. An everyday distinction such as "left" and "right" only makes sense to the robot if it can be expressed in numerical terms. The distinction between past and future also results from the calculation of the ratio of measurable points in time. That humans can experience a future as (respectively) near or far is made possible here only by the expression of a calculable time span (see also [31]). What robots are not able to do is to take the position of another agent [32]. Furthermore, the work of Lindemann and Matsuzaki [32] shows that safety is also very significant for robotics developers. On the one hand, it is important that the safety of people interacting with the robot remains assured. On the other hand, it is necessary that the robot is controllable and reliable, so that technicians can understand how the robot completes its actions [32] (see also [59]). To increase trust in robots, consideration is also being given to integrating risk calculation with regard to decision-making. If robots are able to calculate and assess risks, then people interacting with the robot can also have greater trust [7].

## 3 METHODS

We conducted two workshops, one oriented to the idea of an online Design Fiction and another to the idea of a Zine-Making, with ten experts from Austria, Germany and Japan, in order to be able to have a more complex vision of future reality with the help of the different expert perspectives (see also [65]).

The workshops involved experts from philosophy, sociology, social policy, law, nursing, and HCI, all of whom have experience with robotics and technical systems in various health-related fields. The goal of the Design Fiction workshop was to demonstrate how science fiction narratives might be applied to design perspectives and what different assumptions regarding robots in care exist in Germany and Japan. Design Fiction is a speculative design practice, helping its users to imagine future design scenarios and ways of using technological systems and reflecting on them against the background of socio-cultural aspects [4,15,50]. The Zine-Making workshop, which we see as a form of cultural probe [9,11,33,49,64] was helpful in visualizing future practices as part of the design process with the help of images. Zine-Making is a form of knowledge sharing, which is based on visualizations and various text formats. It has different roots, which are based on a do-it-yourself approach that can also enable marginalized groups to bring their voice into the discourse [16,57,58]. What emerged in our workshops, as we will see below, was how the relevance of autonomy and safety, already a theme in the literature, had to be understood as an outcome of tripartite interactions between caregivers, care-receivers, and robots.

## 4 RESULTS

We refer above to the research aim of assessing the conceptual contributions that resulted from workshops with experts, in relation to the known tension between autonomy and safety in care situations. In doing so, we want to contribute to how to think about the use of robotic systems in care for older persons. We identify several aspects regarding decision-making, which we will address here, drawing on the expertise made visible in the workshops. In particular, it became clear that the autonomy of the robot to make decisions is related to how it is integrated in social relationships:

> "I think autonomy really depends on which context you are using the robot. [...] We are quite often working in hybrid teams. So, there will be a caretaker and a robot and they're both together giving a course or exercise for residents or the group of residents. And then autonomy is not that important and might even be a problem because the caretaker is supposed to have control over the robot." – **Male German Expert: HCI (Design Fiction, Group 1)**

Caregivers are in a position where they want to be sure what will happen in any given situation. It would be better, it is felt, if the robot functioned in a standardized way and had less autonomy. That way, everyone involved would know what the robot is doing and how to respond to it.

In addition, caregivers are the ones who influence the care situations. Their skills and expertise determine in practice the limits of how the robot can be used. The assumption may be that caregivers have a high autonomy in care situations (caregiver, care-receiver, robot) due to their position in the social structure of a care facility and therefore also bear the responsibility to take care of and use the technical artifacts. Again, this would be a basis for negotiation, but one that grants caregivers certain options:

> "But they should use the robot in such a way that they, the care workers, are enabled to participate. That is, with those in need of care. Or with each other, whatever. And not be used as pure human robots, that have to do everything as quickly as possible and as economically as possible and so on." – **Male German Expert: Philosophy (Zine-Making)**

In balancing autonomy and safety, there is a concern that robots will no longer be used at all if the main focus is to maintain the safety of humans. According to the **male German HCI expert**: "There won't be a one hundred percent safety and if we want robots to work autonomously then we need to give them a certain allowance to do something". The logic behind this is that interaction with other people is always associated with a certain risk and that people also have to accept a degree of risk in their lives. The question would therefore also be that if people can autonomously decide that they wish to interact with a robot, even if this is risky, whether it should not nevertheless be permitted. However, these considerations leave liability issues unresolved because it is not clear who is liable for the robot or how it will be legally handled if the robot injures a person:

> "Many international bodies have been discussing introducing many laws for artificial intelligence as well as robotics. So, I think there will be actually quite in the near future to be established, what for example the vision of artificial intelligence means legally speaking or what kind of autonomy they are entitled to. I think all of these quite of confusing areas might be fixed maybe in two or three years maximum and so I think in ten years there will be quite a good robust legal background as far as I know. Well, as far as the situation allows us to tell." – **Female Japanese Expert: Law (Design Fiction, Group 1)**

At a meso level, it is also important that there is certainty of expectation when it comes to future references. The expert points out that in the near future, criteria will be in place at the legal level if robots or AI are to be used. This would provide more certainty with regard to the autonomy of robots and how they can be used in different areas. This would also make robots less likely to be experienced as an unsafe black box, because laws can determine their capabilities and thus their autonomy. However, this statement also makes it clear that the experts are not referring to a distant future, but would rather like to focus on statements that are in the near future and that can be anticipated:

> "And it's not in Japan but for example in Denmark there are kind of institutionalized certificates for those who use animal robots in care facilities. It's so to say acknowledged as a kind of skill, ability to use a robot for care purposes. The role is kind of on an institutional level." – **Male Japanese Expert: Sociology (Design Fiction Group A)**

In particular, there may be sociocultural differences that mediate the acceptance of robots. This may also be present at the legal level when the focus is turned to institutionalization. It can be institutionalized at the organizational level that a certain handling is required when using robots. This can then be integrated into the training of nurses. The nurses receive a certificate that authorizes them to use robots in nursing situations for nursing purposes. This may also be something that changes the view of the nursing profession. It may be that in the future, nurses will be expected to have additional skills related to the use of technical artifacts. This might also be the use of a robot, for example.

Finally, criticism of the workshop format becomes noticeable:

> "The focus is not so: 'Defining care support by myself'. It's more defining it among the community people, yeah? And I meant not that not scientists should define how a community robot is being used. But the community should negotiate and also have the socio-technical environment to negotiate and to interpret what we want to use a robot for. And for what not maybe" – **Female German Expert B: HCI (Design Fiction, Group A)**

The expert said that, in her view, the workshop is about scientists discussing together what a robot can be used for. For this reason, the perspective of the members of a community is emphasized here. This conveys the orientation of the workshop. The aim is not to formulate criteria for how such a robot could function so that it can be used in a community. Rather, it should be about how means can be defined for all stakeholders and robots to be involved in these heteromated decision-making processes.

## 5 DISCUSSION AND CONCLUSION

Although we could not discover any clear differences between the German or Japanese perspectives, the experts' discussions have shown that the practices of the individual actors must be thought of in terms of group structures. There are complex overall actions in which both the robots and the humans can be involved. This is related to the concept of heteromation [14], where individual actions are performed technically and others are performed humanly. In this context, it is necessary to work out how heteromation can be designed so that autonomy and safety or decision-making and trust are adequately addressed and vulnerabilities that are experienced as inappropriate are not continued. In our view, this is a question of how autonomy and safety are experienced and it should be left to relevant stakeholders to decide what options they wish to take into consideration. However, it is becoming clear that autonomy and safety can be addressed through both practices and technologies, and the socio-technical structure in which this occurs is a complex one. This will, in future, become a more complex and consequential question as, for instance, when large language models like Chat GPT are integrated into robots in instructional or advice giving, or other supportive roles. Critically, in robotics in particular, the focus is sometimes on automating care tasks, and the use of care is sometimes supposed to be mainly about making care work cheaper [52]. In our opinion, this perspective precisely lacks the awareness of interactional complexity that is revealed through our workshop activities. Nursing is a complex process in which body work, emotional work, administrative tasks, and information management, among others, must be coordinated. How exactly robots are to be used in these areas requires some reflexive work and better insight into different areas of care [13,30,34,39,40,51]. We hope to take a first step in a promising direction with the concept of heteromated group-robot interaction [14,46].